%% file: paper.tex
\documentclass[10pt,technote,compsoc]{IEEEtran}

\usepackage{xfrac}
\usepackage{booktabs} 
\usepackage{comment}

\usepackage{mathptmx} 
\usepackage{fancyhdr}
\usepackage[normalem]{ulem}
\usepackage[hyphens]{url}
\usepackage[nocompress,nosort]{cite}
\usepackage[final]{microtype}
\usepackage{flushend}
\usepackage{kotex}
\usepackage{subcaption}

\ifCLASSOPTIONcompsoc
  \usepackage[nocompress]{cite}
\else
  \usepackage{cite}
\fi
\ifCLASSINFOpdf
\else
\fi

\begin{document}

\title{LP5X-PIM Sim: A High-Fidelity HW/SW Integrated Simulator for LPDDR5X-PIM}
%

%
%
%
%

\author{
  SangHoon Cha, 
  Jaewan Choi, 
  Byeongho Kim, 
  Yoonah Paik, 
  Sukhan Lee and Kyomin Sohn
 \IEEEcompsocitemizethanks{\IEEEcompsocthanksitem
 All authors are with Samsung Electronics, South Korea.\protect\\  Corresponding E-mail: \{s.h.cha, sh1026.lee\}@samsung.com\protect
 }%
}%

\input{0_abstract}
\maketitle
\IEEEdisplaynontitleabstractindextext

%
\IEEEpeerreviewmaketitle

\input{1_introduction}

\input{2_pim_simulator}
\input{4_evaluation}
\input{5_conclusion}
\newpage

\bibliographystyle{IEEEtran}                                                    

\end{document}

%% file: 0_abstract.tex
\IEEEtitleabstractindextext{%
\begin{abstract}

This tech note describes the architecture and execution results of the LPDDR5X-PIM simulator, developed by Samsung Electronics. Based on the latest research and internal specifications, the simulator provides a high-fidelity model of both the hardware data paths and the software control layers of the LPDDR5X-PIM block. This integrated hardware-software simulation approach enables precise evaluation of system performance and energy efficiency while maximizing PIM resource utilization. We have refined existing simulation frameworks to align with actual hardware implementation, ensuring consistent behavioral accuracy. Further technical details regarding the specific architecture and circuit design of the LPDDR5X-PIM will be disclosed in future publications.

\end{abstract}
}

%% file: 1_introduction.tex
\section{Introduction}
\label{sec:introduction}

Although various PIM simulators~\cite{upmem-sim,sait-pimsimulator,attacc} have been developed recently, most are limited to generic HBM or DDR architectures and suffer from a critical limitation in that they fail to accurately reflect the architecture of commercial LPDDR5X with PIM technology (hereafter referred to as LP5X-PIM) and the standard timing for LPDDR5X. To address these limitations, this paper introduces LP5X-PIM Sim, a dedicated simulator capable of precisely modeling and evaluating Samsung's LP5X-PIM architecture~\cite{lp5x-pim} and its software control layer. The core objective of this simulator is to accurately predict the performance of LP5X-PIM technology and to derive optimization strategies. To achieve this, it is built upon DRAMSim3~\cite{dramsim3} and Ramulator~\cite{ramulator}, which are widely used memory simulators in academia, and is implemented as a cycle-accurate simulator by strictly reflecting Samsung's proprietary PIM technical specifications and LPDDR5X standard operational timings. In particular, it is designed so that a hardware model detailing the datapaths between LP5X-PIM blocks and modules organically interacts with a software model that maximizes resource utilization through efficient PIM kernel operations, allowing for a highly reliable evaluation of system performance.





We propose the first simulation model that strictly captures of Samsung's LP5X-PIM, fully complying with LPDDR5X standard timings to enable precise cycle-level performance prediction.
Also, we model the critical interactions between the software control layer and PIM hardware. By explicitly simulating the behavior of PIM kernel, we enable the exploration of scheduling optimization strategies that maximize actual hardware utilization.

%% file: 2_pim_simulator.tex
\section{Integrated Simulator Design and Implementation}
\label{sec:lp5xpim-sim}

\begin{figure}[!b]
  \center
  \includegraphics[width=0.7\textwidth]{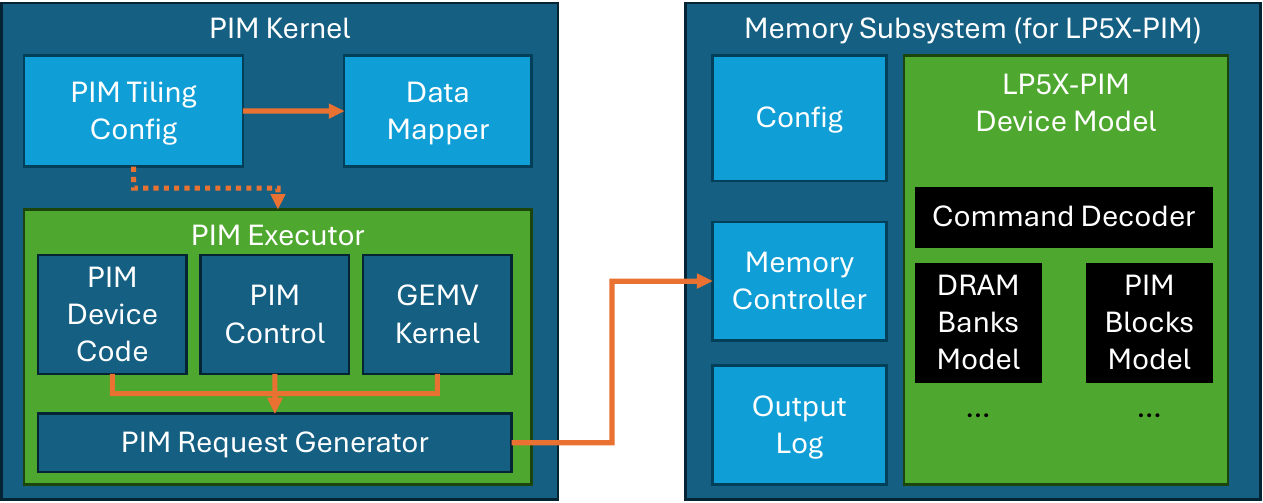}
  \caption{
    Overall block diagram of the LP5X-PIM simulator. The simulator consists of the LP5X-PIM hardware model and the PIM Kernel software model. PIM requests (memory requests) generated by the PIM Kernel are issued to the memory controller, which manages the PIM operations and control logic within the LP5X-PIM device.
  }
  \label{fig:sim_overview}
  \vspace{-0.09in}
\end{figure}

LP5X-PIM Sim is designed to faithfully replicate the behavior of a real-world system through the organic integration of two primary modules: the LP5X-PIM module for hardware modeling and the PIM Kernel module for software control (Figure~\ref{fig:sim_overview}). By precisely simulating cycle-by-cycle interactions between these layers, the simulator provides a highly reliable environment for evaluating hardware-software co-optimization strategies.

\subsection{Hardware Modeling: Memory Controller and LP5X-PIM Device}

The hardware architecture of the simulator is broadly divided into the Memory Controller and the LP5X-PIM Device.  
\\
\textbf{Memory Controller:} It analyzes host memory requests and schedules them to maximize processing throughput while strictly adhering to LPDDR5X standard timing constraints.  
\\
\textbf{LP5X-PIM Device:} This module integrates PIM acceleration logic into a conventional LPDDR5X module to handle both data storage and computation. Each PIM block is deployed in a 1-to-1 mapping with a corresponding DRAM bank, allowing direct data transfer through internal compute units and specialized PIM registers. This structural characteristic fundamentally circumvents bandwidth bottlenecks typically caused by off-chip data movement.

\subsection{Software Modeling: PIM Kernel Layer}
\begin{figure}[!tb]
  \center
  \includegraphics[width=0.8\textwidth]{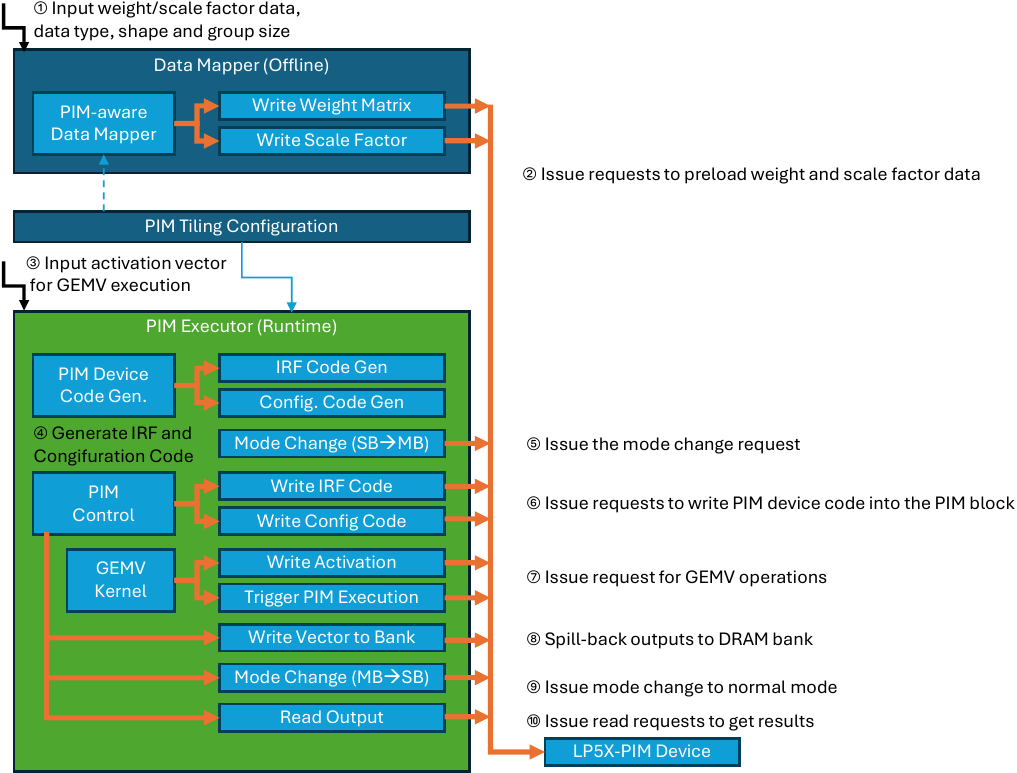}
  \caption{
    Execution flow of the PIM kernel. The flow is divided into two main components: the DataMapper, which operates offline, and the PIM Executor, which operates at runtime. Both components refer to the PIM tiling configuration. Based on this configuration, the DataMapper performs PIM-aware data placement, while the PIM Executor executes the PIM operations.
  }
  \label{fig:lp5xpimkernel}
  \vspace{-0.09in}
\end{figure}

The PIM Kernel serves as the software modeling layer that manages LP5X-PIM hardware resources and mediates application-level computations (Figure~\ref{fig:lp5xpimkernel}). It is composed of the Data Mapper for offline data placement and the PIM Executor for runtime control.  
\\
\textbf{Data Mapper (Offline):} It receives target parameters such as the weight matrix and data types to structure data into "PIM Tiles". By referencing a predefined PIM Tile Configuration, the Data Mapper generates an optimal memory layout and preloads it into DRAM banks, eliminating the latency of data rearrangement during runtime.  
\\
\textbf{PIM Executor (Runtime):} It manages active computations through three core sub-components. \\
1) PIM Device Code Gen: Dynamically synthesizes optimized PIM instructions (IRF code) and hardware configuration code based on matrix shapes and data types. \\
2) PIM Control: Manages system-wide logic, including mode transitions between Single-Bank (SB) and Multi-Bank (MB) modes. SB mode is used for standard DRAM operations, while MB mode enables parallel PIM execution across multiple banks.
\\
3)  GEMV Kernel: Executes General Matrix-Vector Multiplication on a per-tile basis using the specialized PIM ISA and manages pipeline flush-out operations.  

\subsection{Address Mapping and Tiling Strategies}
\label{sec:lp5xpim-addrtile}

\begin{figure*}[!tb]
  \center
  \includegraphics[width=0.9\textwidth]{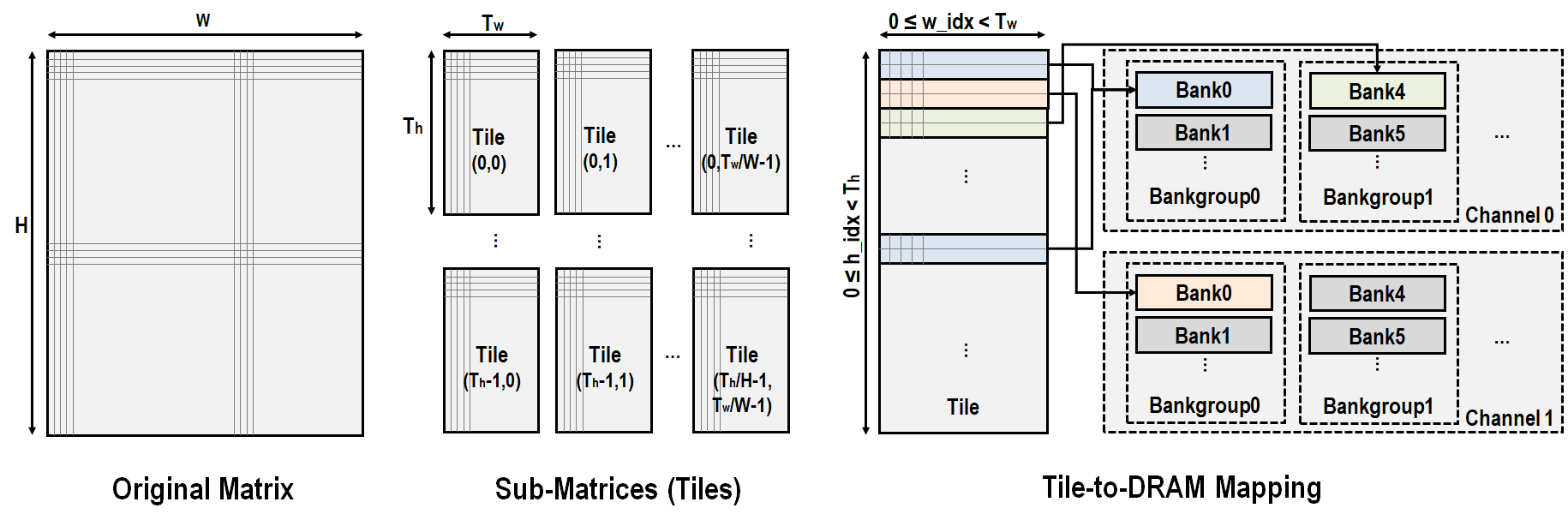}
  \caption{
     An example of the PIM tiling configuration. The original matrix of size $H \times W$ is partitioned into multiple sub-matrices of size $T_h \times T_w$, which are referred to as PIM tiles. When physically mapping these PIM tiles to DRAM, they are placed into DRAM banks in a manner that maximizes the parallel execution of multiple PIM blocks.
  }
  \label{fig:pim_tile}
  \vspace{-0.09in}
\end{figure*}

The simulator employs sophisticated 2D-based address mapping and tiling techniques to minimize data movement and control overhead (Figure~\ref{fig:pim_tile}). 
Fundamentally, the tile size is constrained by the capacities of the PIM block's input/output register files and the data precision.
\\
\textbf{Vertical Mapping:} Rows within a tile are sequentially interleaved across DRAM hierarchies (Channel, Rank, Bank Group, Bank) to maximize bank-level parallelism and external memory bandwidth utilization.\\
\textbf{Horizontal Mapping:} Adjacent sub-matrices are allocated within the same bank to drastically improve the row buffer hit rate during sequential access.\\  
\textbf{Reshape Optimization:} To address hardware under-utilization, the kernel applies a column-based partitioning technique. By distributing data across both row ($H$) and column ($W$) dimensions, the simulator ensures near-100\% intra-PIM efficiency, which is particularly effective for small-scale matrices that would otherwise fail to utilize all available PIM blocks. 


%% file: 4_evaluation.tex
\section{Evaluation}
\label{sec:evaluation}


This section evaluates the performance and efficiency of the LP5X-PIM architecture and the PIM Kernel based on comprehensive simulations using our LP5X-PIM Sim. The evaluation is conducted across five primary dimensions: (1) baseline GEMV acceleration without memory-fence overhead, (2) acceleration in a realistic system constrained by memory fences.

The reference memory system consists of LPDDR5X-9600 and strictly complies with JEDEC~\cite{jedec-lp5x}-based timing specifications. All experiments operate with four DRAM channels. Through LP5X-PIM Sim, we meticulously captured the precise timing, and data movement of both standard memory operations and internal PIM executions.

\begin{figure*}[!t]
     \centering
     \begin{subfigure}{0.9\textwidth}
         \centering
         \includegraphics[width=\linewidth]{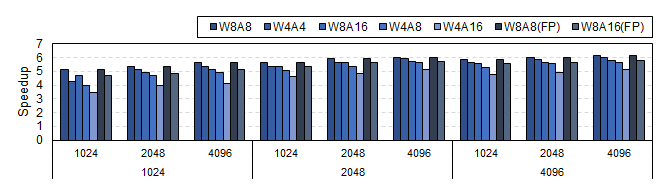}
           \vspace{-0.29in}
         \caption{Without memory-fence latency}
         \label{fig:perf_no_fence}
     \end{subfigure}
     
     \vspace{0.2cm}
     
     \begin{subfigure}{0.9\textwidth}
         \centering
         \includegraphics[width=\linewidth]{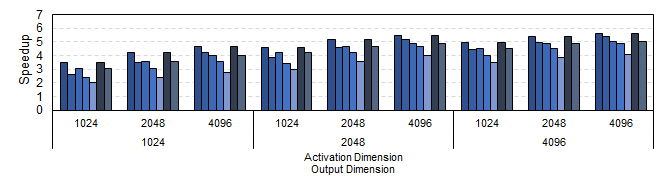}
         \vspace{-0.29in}
         \caption{With a memory-fence latency of 150 ns}
         \label{fig:perf_150_fence}
     \end{subfigure}
     
     \caption{ Performance evaluation of LP5X-PIM for GEMV acceleration across various data shapes and types. The subfigures show the speedup (a) without a memory fence and (b) with a 150 ns memory fence. The speedup is normalized to the sequential weight read latency of a non-PIM baseline system configured with four DRAM channels.}
     \label{fig:overall_performance}
  \vspace{-0.09in}
\end{figure*} 

\subsection{GEMV Acceleration}
Figure~\ref{fig:perf_no_fence} illustrates the PIM speedup ratio compared to a baseline system (non-PIM, sequential read) in an environment where memory request ordering is guaranteed. The evaluation utilized various integer (W8A8, W4A4, W8A16, W4A8, W4A16) and floating-point (W8A8, W8A16) data types across expanding GEMV dimensions, with the top and bottom panels representing the activation dimension and the output vector dimension, respectively.

Regardless of the data type, the acceleration performance consistently improves as the dimensions of the matrix increase. This trend occurs because larger computation workloads effectively amortize the fixed latency of PIM control operations, such as mode transitions, pipeline flush-outs, and accumulation register-to-DRAM data movements.

When categorized by weight tile shape at a baseline dimension of 4096, configurations with larger tile shapes (e.g., W8A8, W4A4, and W8A8 FP) achieved an optimal speedup of 6.0$\times$  to 6.2$\times$. Configurations with smaller tile shapes (e.g., W8A16, W4A16, and W8A16 FP) yielded slightly lower speedups ranging from 5.7$\times$  to 5.8$\times$ . This variance is primarily due to the increased frequency of writing input vectors to the Source Register File (SRF) per tile, which introduces additional overhead for smaller tile shapes.


\subsection{Impact of Memory-Fence Overhead}
In the system, the on-chip bus and the memory controller may alter the order of memory requests or DRAM commands. To prevent this, memory-fence instructions can be used between successive tiles to strictly guarantee inter-tile execution order. However, because memory fences introduce additional overhead, they can negatively impact performance improvements. Although memory-fence latency varies depending on the host processor type and system load, we conducted simulations setting the static memory-fence latency to 150 ns, a representative value for high-performance mobile application processors confirmed through empirical testing.

Figure~\ref{fig:perf_150_fence} presents the simulation results, including the memory-fence latency. While the architecture still maintained a robust speedup of over 5.0$\times$ for the 4096 dimension across most configurations, the W4A16 format (which possesses the smallest tile shape) saw its speedup drop to 4.1$\times$. Because smaller tile shapes generate a higher total number of tiles, they invoke memory fences between tiles more frequently, leading to a proportionally larger performance penalty. However, similar to the fence-less scenario, larger computation dimensions successfully amortize this fixed fence overhead, maximizing the net PIM acceleration.

\subsection{Effectiveness of Reshape Optimization}
We further evaluated the impact of the Reshape Optimization strategy mentioned in Section~\ref{sec:lp5xpim-addrtile}. For small matrix dimensions (e.g., cases where $W < 2048$), the default row-only partitioning often leads to idle PIM blocks. By enabling column-based reshape partitioning, the PIM Kernel effectively activated more PIM blocks and maximized register utilization. Our simulation results demonstrate that this software-level optimization provides an additional performance gain of up to 1.65$\times$ compared to the non-reshaped baseline, validating the simulator's capability in exploring software stack optimizations.

%% file: 5_conclusion.tex
\section{CONCLUSION}



In this paper, we proposed LP5X-PIM Sim, a cycle-accurate simulator designed to rigorously model and evaluate Samsung's LPDDR5X-PIM architecture. To address the escalating data movement bottleneck in modern computing, our simulator provides a highly reliable evaluation environment by integrating our hardware logic and datapath with the software control layer via the PIM kernel. 
Leveraging this tool, we demonstrated that LP5X-PIM technology can achieve up to a 6.2$\times$ performance acceleration in GEMV operations. Furthermore, we verified that tile-level optimization, such as Reshape, yielded an additional 1.65$\times$ gain. LP5X-PIM Sim serves as a robust framework for exploring PIM architectures and software stacks. While this paper focuses on the simulation framework and its evaluation, more detailed information regarding the specific architecture and circuit-level implementations of LPDDR5X-PIM will be disclosed in our future publications.